\documentclass[11pt,a4paper]{article}

\usepackage[utf8]{inputenc}
\usepackage[T1]{fontenc}
\usepackage[margin=2.5cm]{geometry}
\usepackage{graphicx}
\usepackage{multirow}%
\usepackage{amsmath,amssymb,amsfonts}%
\usepackage{amsthm}%
\usepackage{mathrsfs}%
\usepackage[title]{appendix}%
\usepackage{booktabs}%
\usepackage{dcolumn}%
\usepackage{bm}%
\usepackage{authblk}
\usepackage[font=small,labelfont=bf,labelsep=period]{caption}
\usepackage[colorlinks=true, allcolors=blue]{hyperref}

\title{Investigation on Thermal Stability of Epi-GeSn Directly on Si (001)}

\author[1]{Tharundev VV}
\author[1]{Dushyant Singh}
\author[1]{Vigneshswari SR}
\author[1]{Subha Maity}
\author[1]{Prince Rathor}
\author[1,*]{Krista R. Khiangte}

\affil[1]{Department of Physics, Indian Institute of Technology Gandhinagar, Gandhinagar, India}
\affil[*]{Corresponding author: \href{mailto:Krista.khiangte@iitgn.ac.in}{Krista.khiangte@iitgn.ac.in}}

\date{}

\begin{document}

\maketitle

\begin{abstract}
	In this paper we investigate the structural properties and thermal stability of metastable Ge$_{1-x}$Sn$_x$ epilayers ($\sim$~4.5 at.\% Sn) grown directly on Si(001) substrates by high-resolution X-ray diffraction (HRXRD) and Raman spectroscopy.  Our results demonstrate that both 70 and 140~nm films are fully relaxed, with the thicker film exhibiting a small tensile strain (+0.09\%). The thermal stability of the heterostructures was evaluated up to 813~K using temperature-dependent Raman spectroscopy and scanning electron microscopy. The surface-weighted Raman response (532~nm excitation) and surface morphology of the 70~nm film remain highly stable. In contrast, the 140~nm film undergoes severe degradation, characterized by significant attenuation and broadening of the Ge-Ge longitudinal optical phonon mode, alongside enhanced surface Sn segregation. These findings demonstrate that the thermal stability of direct GeSn/Si heterostructures is governed by the initial defect structure established during epitaxial growth, providing key design rules for understanding the fabrication of GeSn-based optoelectronic devices.
\end{abstract}


\section{Introduction}

Silicon photonics has become a key technology for optical interconnects, data communications and integrated sensing. Nevertheless, the absence of an efficient, monolithically integrated group-IV light source remains one of the principal limitations of silicon photonic platforms \cite{Thomson2016,Liang2010}. Because both Si and Ge possess indirect bandgaps, efficient radiative recombination is intrinsically limited. Although heterogeneous integration of III-V semiconductors provides an effective solution, it increases fabrication complexity, processing cost and thermal-budget requirements, thereby reducing compatibility with conventional complementary metal-oxide-semiconductor (CMOS) manufacturing \cite{Liang2010}. The development of a monolithic group-IV light source is therefore regarded as a major objective for next-generation silicon photonics \cite{Soref2010,shekhar2024}.

Among the candidate materials, Ge$_{1-x}$Sn$_x$ has emerged as the most promising group-IV alloy owing to its tunable electronic band structure. Incorporation of Sn progressively lowers the energy of the direct $\Gamma$ valley relative to the indirect $L$ valley, resulting in an indirect-to-direct bandgap transition at Sn concentrations of approximately 6-8 at.\% \cite{khiangte2018,reboud2024,zheng2018, Singh2025Raman}. This prediction has been confirmed experimentally through the demonstration of optically pumped GeSn lasers grown on Si substrates \cite{Wirths2015}, followed by rapid advances in operating temperature, emission wavelength and device performance \cite{margetis2017,Elbaz2020}. In parallel, GeSn-based photodetectors and light-emitting diodes have further established the potential of this material system for silicon-integrated mid-infrared photonics \cite{zheng2024,atalla2024}.

Most high-performance GeSn epilayers are grown on relaxed Ge virtual substrates, which reduce the lattice mismatch between GeSn and the underlying template and thereby facilitate high-quality epitaxial growth \cite{giunto2024,reboud2024}. However, the use of thick Ge buffers increases the overall heterostructure thickness, requires additional high-temperature processing steps and complicates subsequent device fabrication. Direct epitaxy of GeSn on Si(001) provides an attractive alternative by eliminating the Ge virtual substrate, reducing the cumulative thermal budget and simplifying integration \cite{jiang2025}. These advantages are particularly attractive for back-end-of-line compatible photonic devices \cite{reboud2024}. The larger lattice mismatch between GeSn and Si, however, introduces a substantially different strain-relaxation behaviour that promotes defect formation and modifies the structural evolution of the epilayer during growth and subsequent thermal processing \cite{jiang2025,cai2022}. Epitaxial lateral overgrowth of Ge on dielectric-masked Si has recently emerged as a promising route to high-quality single-crystalline epilayers on Si~\cite{Zhang2023,Zhang2024}.

Therefore, thermal stability represents a key challenge for the practical implementation of GeSn devices. Post-growth fabrication processes, including dopant activation and ohmic contact formation, commonly involve rapid thermal annealing at temperatures between 673 and 873~K~\cite{moutanabbir2021,von2020,Wirths2015}. Such thermal treatments can activate Sn diffusion, segregation and phase separation, limiting the structural stability of the metastable GeSn alloy. Under these conditions, the metastable nature of GeSn promotes several competing degradation mechanisms, including Sn segregation, strain relaxation through dislocation formation and the evolution of point defects \cite{jung2025,giunto2024}, that influence the electronic and optical properties of the material \cite{zhou2025,baral2026}. Although these processes have been investigated for GeSn grown on Ge virtual substrates, considerably less is known about their behaviour in GeSn layers grown directly on Si, where the initial strain state and defect structure differ substantially. Understanding the thermal evolution of direct GeSn/Si heterostructures is therefore essential for the realization of reliable group-IV photonic devices.

GeSn alloys containing approximately 5 at.\% Sn provide an ideal model system for such investigations. At this composition the alloy remains indirect bandgap, while the Sn concentration is still well above its equilibrium solubility in Ge. Consequently, the effects of alloy metastability, defect evolution and thermally activated structural degradation can be examined without the severe phase separation that frequently accompanies higher Sn concentrations \cite{xu2025,giunto2024}. Furthermore, the vibrational properties of low-Sn GeSn alloys are well established through the characteristic Ge-Ge, disorder-activated Ge-Ge and Ge-Sn Raman modes \cite{corley2025,vv2026}, enabling Raman spectroscopy to serve as a sensitive probe of local structural evolution.

In this work, we investigate the influence of film thickness on the structural quality and thermal stability of Ge$_{1-x}$Sn$_x$ epilayers containing approximately 4.5 at.\% Sn grown directly on Si(001). High-resolution X-ray diffraction, reciprocal-space mapping and pole-figure analysis are first employed to establish the crystallographic quality, strain state and epitaxial relationship of the as-grown films. The thermal evolution of the epilayers is subsequently examined by temperature-dependent Raman spectroscopy, complemented by two-dimensional Raman mapping and scanning electron microscopy. By correlating the structural, vibrational and morphological characteristics, we demonstrate how the initial crystalline quality governs the thermal response of direct GeSn/Si heterostructures and identify the key mechanisms responsible for thickness-dependent degradation.

\section{Experimental}
Ge$_{1-x}$Sn$_x$ epilayers with thicknesses of 70 and 140 nm were grown directly on Si(001) substrates by molecular beam epitaxy (MBE). Details of the growth procedure are provided elsewhere \cite{khiangte2017}. The structural properties of the GeSn epilayers were investigated by high-resolution X-ray diffraction (HRXRD) and reciprocal-space mapping (RSM) around the symmetric (004) and asymmetric (224) reflections using Cu K$\alpha_1$ radiation ($\lambda = 1.54$~\AA). The in-plane epitaxial relationship and defect evolution were further examined by X-ray pole-figure measurements around the asymmetric (224) reflection. Local vibrational properties and lateral structural uniformity were characterized by room-temperature micro-Raman spectroscopy and two-dimensional Raman mapping using a 532 nm excitation laser over a $20 \times 20~\mu\mathrm{m}^{2}$ area. Temperature-dependent Raman measurements were subsequently performed using the same excitation wavelength over the temperature range of 100-813 K with a step size of 40 K. The surface morphology of the epilayers was examined by plan-view scanning electron microscopy (SEM) operated at an accelerating voltage of 5 kV before and after the temperature-dependent Raman measurements.

\section{Results and Discussion}

\begin{figure}[!htb]
	\centering
	\includegraphics[width=1\linewidth]{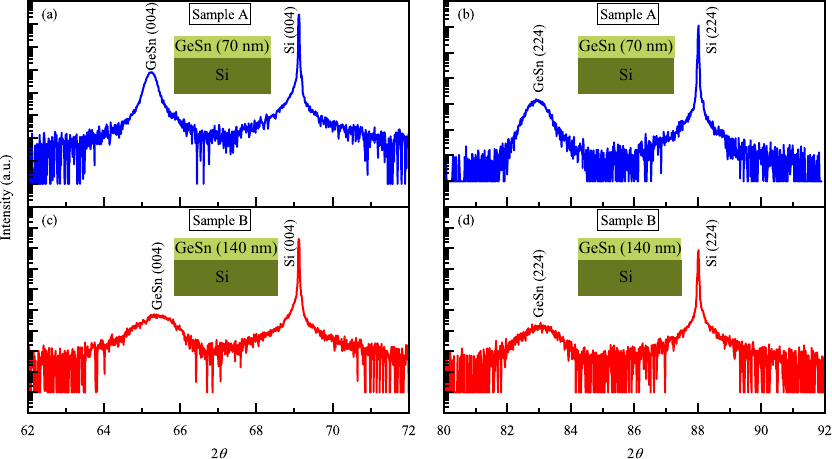}
	\caption{High-resolution X-ray diffraction scans of the as-deposited GeSn thin films on Si(001): (a)~symmetric $\omega$-$2\theta$ (004) and (b)~asymmetric (224) scans of sample~A (70~nm); (c)~symmetric (004) and (d)~asymmetric (224) scans of sample~B (140~nm). Both samples display a distinct GeSn epilayer peak alongside the sharp Si substrate reflection, confirming single-phase epitaxial growth.}
	\label{fig:xrd}
\end{figure}

Figure~\ref{fig:xrd}(a-d) shows the high-resolution symmetric (004) and asymmetric (224) $\omega$-2$\theta$ diffraction patterns of the 70 and 140 nm GeSn epilayers grown on Si(001). In both samples, a distinct GeSn diffraction peak is observed at a lower Bragg angle than the intense Si(004) substrate reflection, reflecting the larger lattice parameter of the GeSn alloy and confirming the epitaxial growth of the films. The diffraction profiles exhibit a pronounced dependence on film thickness. Compared with the 70 nm epilayer, the 140 nm film displays broader and less intense diffraction peaks in both reflections, indicating a reduction in crystalline coherence. As the GeSn epilayer thickens, the large lattice mismatch with the Si substrate is progressively accommodated through defect-mediated strain relaxation, leading to increased mosaicity and a broader distribution of lattice spacings \cite{jiang2025}. Consequently, the diffraction peaks broaden, indicating a progressive deterioration of the crystalline quality with increasing film thickness.

\begin{figure}[!htb]
	\centering
	\includegraphics[width=0.5\linewidth]{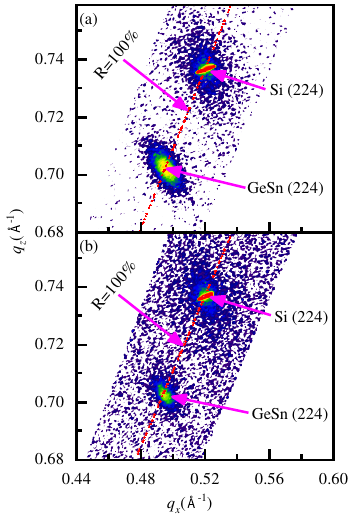}
	\caption{Reciprocal space maps around the asymmetric (224) Bragg reflections for the as-deposited GeSn/Si(001) heterostructures: (a)~sample~A (70~nm) and (b)~sample~B (140~nm).}
	\label{fig:rsm}
\end{figure}

To further evaluate the strain state and alloy composition, reciprocal-space maps (RSMs) were acquired around the symmetric (004) and asymmetric (224) reflections, as shown in Fig.~\ref{fig:rsm}. The symmetric reflection provides the out-of-plane lattice parameter, whereas the asymmetric reflection enables determination of the in-plane lattice parameter, allowing the strain and composition of the GeSn epilayers to be evaluated independently using the elastic theory of tetragonally distorted cubic crystals together with the modified Vegard's law~\cite{dev2020}. The in-plane ($a_{\parallel}$) and out-of-plane ($a_{\perp}$) lattice parameters were extracted from the position of the GeSn reciprocal-lattice point in the (224) RSM~\cite{rathore2021,aubin2017}. The relaxed lattice parameter ($a^{\text{GeSn}}$), Sn content ($x$) and degree of strain relaxation ($R_{\text{GeSn}}$) with respect to the Si(001) substrate were then determined following the procedure described in Refs.~\cite{dev2020,GAYAKWAD2023}. The elastic stiffness constants of the GeSn alloy were obtained by linear interpolation between those of bulk Ge ($C_{11}^{\text{Ge}} = 128.5$~GPa, $C_{12}^{\text{Ge}} = 48.3$~GPa) and bulk $\alpha$-Sn ($C_{11}^{\text{Sn}} = 72.5$~GPa, $C_{12}^{\text{Sn}} = 29.9$~GPa)~\cite{fleischmann2016}. The Sn content was obtained from the relaxed lattice parameter using the modified Vegard's law~\cite{gencarelli2013} with $a^{\text{Ge}} = 5.6578$~\AA\ and $a^{\text{Sn}} = 6.4892$~\AA, and $a^{\text{Si}} = 5.431$~\AA\ was used for the Si substrate. A relaxation value of $R = 100\%$ corresponds to a fully relaxed GeSn epilayer, whereas values exceeding 100\% indicate over-relaxation, for which the in-plane lattice parameter exceeds the relaxed lattice parameter of the alloy. The residual in-plane biaxial strain is subsequently determined from $\varepsilon_{\parallel} = (a_{\parallel}^{\text{GeSn}} - a^{\text{GeSn}}) / a^{\text{GeSn}}$.

\begin{table}[!htb]
	\caption{Structural parameters of the GeSn epilayers obtained from the (004) and (224) RSMs. Uncertainties: $\pm 0.001$~\AA\ (lattice parameters) and $\pm 0.1$~at.\% (Sn content).}
	\label{tab:xrd_parameters}

	\centering
	\small
	\setlength{\tabcolsep}{4.5pt}
	\begin{tabular}{lccccccc}
		\toprule
		Sample & Thickness & $a^{\text{GeSn}}_{\perp}$ (\AA) & $a^{\text{GeSn}}_{\parallel}$ (\AA) & $a^{\text{GeSn}}$ (\AA) & Sn (at.\%) & $\varepsilon_{\parallel}$ (\%) & $R_{\text{GeSn}}$ (\%) \\
		\midrule
		A - as-deposited & 70 nm  & 5.695  & 5.695 & 5.695 & 4.5 & $+$0.00 & 100.0 \\
		B - as-deposited & 140 nm & 5.696  & 5.705 & 5.700 & 5.0 & $+$0.09 & 101.9 \\
		\bottomrule
	\end{tabular}

\end{table}

Applying this analysis to the measured reciprocal-space maps yields the structural parameters summarized in Table~\ref{tab:xrd_parameters}. The 70 nm GeSn epilayer exhibits in-plane and out-of-plane lattice parameters of $a_{\parallel}=5.695$~\AA\ and $a_{\perp}=5.695$~\AA, respectively, whereas the corresponding values for the 140 nm film are $a_{\parallel}=5.705$~\AA\ and $a_{\perp}=5.696$~\AA, respectively. The relaxed lattice parameters are $a_{\mathrm{GeSn}}=5.695$~\AA\ and $5.700$~\AA, corresponding to Sn concentrations of 4.5 and 5.0 at.\% for sample A and sample B, respectively. The calculated relaxation values of 100.0\% and 101.9\% indicate that sample A is completely relaxed while sample B is slightly over-relaxed (tensile strained) at room temperature, with residual in-plane strains of 0.0\% and $+$0.09\%, respectively. Within the experimental resolution of the HRXRD reciprocal-space mapping measurements, no measurable difference between the in-plane and out-of-plane lattice parameters was observed for sample A, indicating that any residual strain is below the detection limit of the measurement. The higher Sn concentration in the thicker film is attributed to strain-relaxation-enhanced Sn incorporation \cite{assali2019,dou2018}; as plastic relaxation progresses near the Si interface, the reduced compressive strain on the growing surface lowers the thermodynamic barrier for Sn incorporation \cite{assali2019}.

Although the average structural parameters remain nearly unchanged, the reciprocal-space intensity distribution evolves markedly with increasing thickness. Compared with the 70 nm epilayer, the reciprocal-lattice point of the 140 nm film is broader and more diffuse, indicating increased lattice distortion, mosaicity and local strain inhomogeneity \cite{assali2019}. This behaviour is fully consistent with the diffraction peak broadening observed in the $\omega$-2$\theta$ scans and reflects the accumulation of extended crystalline defects during thickness-dependent strain relaxation. Additionally, the longer growth duration required for the thicker film increases the cumulative thermal exposure, including radiative heating from the hot effusion cells, which can promote defect propagation and Sn diffusion during epitaxy. Together, these results indicate that increasing the GeSn thickness from 70 to 140 nm primarily degrades the crystalline quality while keeping the average alloy composition and strain state close to the relaxed regime.

To quantitatively investigate these structural broadening mechanisms, reciprocal lattice point (RLP) shape analysis was performed on the asymmetric (224) reciprocal space maps. Following established methodology \cite{heinke1994,dolabella2022}, the full widths at half maximum (FWHMs) of the GeSn RLP were extracted parallel ($\Delta q_{\parallel}$) and perpendicular ($\Delta q_{\perp}$) to the scattering vector $\vec{Q}$. The extracted broadening parameters and their ratios are summarized in Table~\ref{tab:rlp_parameters}. The parallel FWHM $\Delta q_{\parallel}$ is governed by the vertical coherence length (thickness) and strain gradients through the layer thickness, whereas the perpendicular FWHM $\Delta q_{\perp}$ is primarily sensitive to the mosaic tilt distribution (mosaicity) and lateral coherence length \cite{heinke1994}. For the 70 nm film (sample A), the ratio $\Delta q_{\perp}/\Delta q_{\parallel}$ is 1.43, demonstrating that the RLP broadening is dominated by mosaic tilt and indicating substantial angular misorientation (mosaicity) relative to strain gradients. In contrast, for the 140 nm film (sample B), the ratio drops to 0.71, driven by a significant increase in $\Delta q_{\parallel}$ from 0.0056 to 0.0095 \AA$^{-1}$. This marked increase in radial broadening ($\Delta q_{\parallel}$) with thickness reflects the development of a depth-dependent strain relaxation gradient as plastic relaxation progresses via misfit dislocations near the Si interface \cite{heinke1994}. Simultaneously, the transverse broadening $\Delta q_{\perp}$ decreases from 0.0080 to 0.0067 \AA$^{-1}$, suggesting that the relative influence of mosaic tilt becomes less dominant or that the lateral grain size increases in the upper portions of the thicker film \cite{heinke1994}. This transition from a mosaicity-dominated to a strain-gradient-dominated broadening regime provides quantitative evidence of the thickness-dependent defect structural evolution in direct GeSn/Si(001) epitaxy.

\begin{table}[!htb]
	\caption{Reciprocal lattice point (RLP) broadening parameters of the GeSn epilayers extracted from the asymmetric (224) reciprocal space maps. $\Delta q_{\parallel}$ and $\Delta q_{\perp}$ denote the FWHM of the GeSn RLP parallel and perpendicular to the scattering vector, respectively.}
	\label{tab:rlp_parameters}
	\centering
	\small
	\begin{tabular}{lccc}
		\toprule
		Sample & $\Delta q_{\parallel}$ (\AA$^{-1}$) & $\Delta q_{\perp}$ (\AA$^{-1}$) & $\Delta q_{\perp}/\Delta q_{\parallel}$ \\
		\midrule
		A (70 nm) & 0.0056 & 0.0080 & 1.43 \\
		B (140 nm) & 0.0095 & 0.0067 & 0.71 \\
		\bottomrule
	\end{tabular}
\end{table}

\begin{figure}[!htb]
	\centering
	\includegraphics[width=0.9\linewidth]{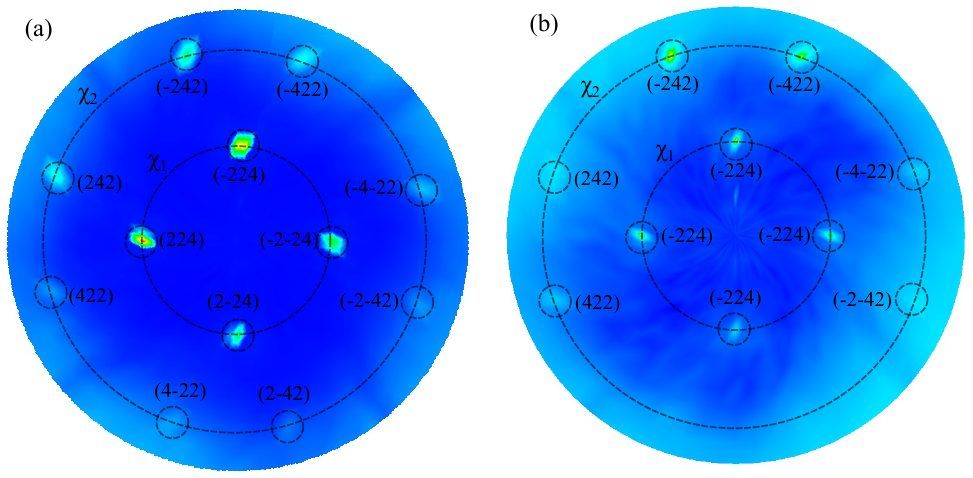}
	\caption{(224) pole figure measurements of the as-deposited GeSn epilayers: (a)~sample~A (70~nm) and (b)~sample~B (140~nm). Sample~A exhibits the four sharp, equal-intensity poles expected for a single-crystalline (001)-oriented cubic film. Sample~B shows weakened and broader poles, indicative of degraded crystallographic coherence arising from planar defect formation during strain relaxation.}
	\label{pf}
\end{figure}

To evaluate the epitaxial relationship and in-plane crystallographic orientation of the group IV epilayers, X-ray pole-figure measurements have been widely used by our group \cite{Krista_Ge2018, Khiangte_2018, singh2025, GAYAKWAD2023}; here, we also perform a pole-figure measurement around the asymmetric (224) reflection for these samples, as shown in Fig.~\ref{pf}(a,b). The 70 nm GeSn film exhibits four intense diffraction poles arranged with four-fold rotational symmetry, characteristic of a (001)-oriented cubic crystal \cite{dev2020}. The measured pole positions coincide with those expected for the {224} family of planes, confirming the cube-on-cube epitaxial relationship between the GeSn epilayer and the Si(001) substrate. No additional diffraction poles are observed within the measured angular range, indicating the absence of secondary crystallographic orientations or twin domains within the sensitivity of the measurement \cite{singh2025}. Together with the low diffuse background intensity, these observations indicate excellent epitaxial quality and a high degree of in-plane crystalline coherence.

In contrast, the pole figure of the 140 nm GeSn film exhibits a pronounced reduction in diffraction intensity accompanied by enhanced diffuse background scattering. Although the four-fold rotational symmetry is preserved, the diffraction poles become broader and less distinct, indicating a reduction in long-range crystalline coherence with increasing film thickness. Because the reciprocal-space mapping analysis shows that both epilayers are fully relaxed (Table~\ref{tab:xrd_parameters}), these differences cannot be attributed to changes in the average strain state. Instead, the diffuse scattering is consistent with increased mosaicity and local lattice distortion arising from defect-mediated strain relaxation during growth. As the epilayer thickness increases, the accumulated lattice mismatch is accommodated through the progressive generation of extended crystalline defects, resulting in a broader distribution of crystallographic orientations and a corresponding reduction in coherent diffraction intensity.

The pole-figure measurements therefore provide a structural picture that is fully consistent with the HRXRD and reciprocal-space mapping analyses. Although both GeSn epilayers retain the same cube-on-cube epitaxial relationship with the Si(001) substrate and exhibit nearly identical alloy compositions and relaxation states, increasing the film thickness from 70 to 140 nm leads to a progressive deterioration of the crystalline quality. The sharper diffraction poles of the thinner epilayer reflect superior crystalline coherence, whereas the broader and more diffuse poles of the thicker film indicate increased mosaicity associated with defect accumulation during strain relaxation.

\begin{figure}[!htb]
	\centering
	\includegraphics[width=1\linewidth]{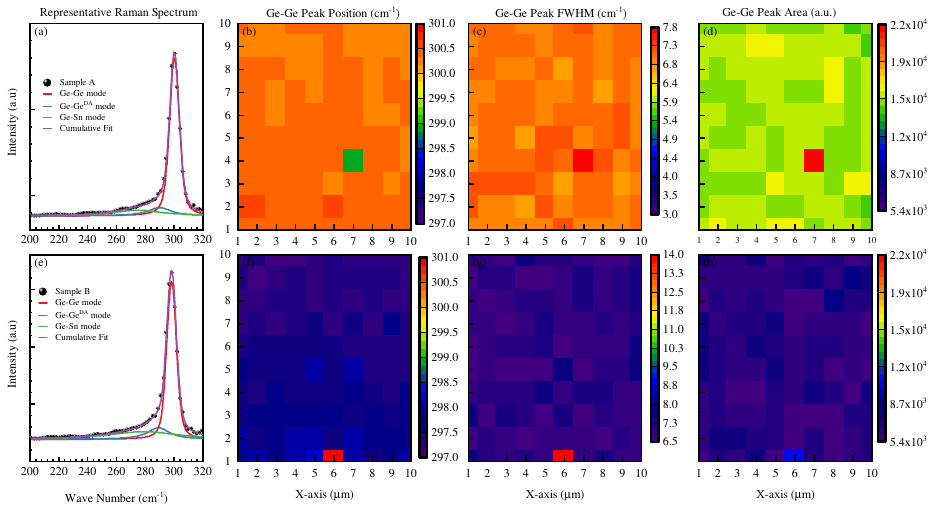}
	\caption{Representative Raman spectra and two-dimensional Raman mapping of the as-grown GeSn/Si(001) samples prior to temperature-dependent Raman measurements. (a, e) Representative Raman spectra with deconvoluted Ge-Ge, disorder-activated Ge-Ge (\mbox{Ge-Ge$_{\mathrm{DA}}$}), and Ge-Sn vibrational modes together with the cumulative fit for sample A and sample B, respectively. (b, f) Ge-Ge peak position maps. (c, g) Ge-Ge peak full width at half maximum (FWHM) maps. (d, h) Ge-Ge peak integrated intensity maps acquired over a $20 \times 20~\mu\text{m}^{2}$ area, demonstrating the spatial homogeneity of the Raman characteristics before thermal cycling.}
	\label{fig:raman_2d}
\end{figure}

To investigate the local vibrational properties and lateral structural uniformity of the as-grown GeSn epilayers, room-temperature micro-Raman spectroscopy together with two-dimensional (2D) Raman mapping was performed using a 532 nm excitation laser. Representative Raman spectra and the corresponding spatial maps of the Ge-Ge longitudinal optical (LO) phonon mode acquired over a $20 \times 20~\mu\mathrm{m}^{2}$ area (consisting of a $10 \times 10$ grid with a step size of $2~\mu\mathrm{m}$) are presented in Fig.~\ref{fig:raman_2d}(a-h). The spectra of both sample A (70 nm) and sample B (140 nm) were fitted using Voigt functions to resolve the characteristic vibrational modes of the GeSn alloy, namely the Ge-Ge LO mode, the disorder-activated Ge-Ge (Ge-Ge$^{\mathrm{DA}}$) mode and the Ge-Sn mode (Fig.~\ref{fig:raman_2d}(a,e)) \cite{liu2020,vv2026}.

Reciprocal-space mapping established that both epilayers are fully relaxed and possess comparable average lattice parameters and Sn compositions (Table~\ref{tab:xrd_parameters}). The Raman maps provide complementary information by probing the local vibrational response near the film surface. For 532 nm excitation, the optical penetration depth in Ge-rich alloys is highly surface-weighted, probing only the near-surface tens of nanometers (9 nm) \cite{corley2025}. Because both films are significantly thicker than this probe depth, the Raman measurements are sensitive primarily to surface-related features. This surface-weighted nature supports linking the observed Raman degradation to the surface precipitates and pits, but also warrants caution against over-interpreting the bulk crystallinity of the deeper regions of the epilayers from Raman analysis alone. Sample A exhibits a mean Ge-Ge phonon position of $300.35~\mathrm{cm}^{-1}$ with a narrow spatial variation of $0.16~\mathrm{cm}^{-1}$, whereas sample B displays a redshifted peak position of $297.55~\mathrm{cm}^{-1}$ together with a larger spatial variation of $0.42~\mathrm{cm}^{-1}$ (Fig.~\ref{fig:raman_2d}(b,f)). The observed Ge-Ge phonon redshift of $2.8~\mathrm{cm}^{-1}$ in sample B relative to sample A reflects the combined influence of its slightly higher Sn concentration ($5.0\%$ vs $4.5\%$), its residual in-plane tensile strain ($+0.09\%$), and the greater degree of crystalline disorder and defect-induced phonon confinement associated with the thicker epilayer.

The structural differences revealed by HRXRD and reciprocal-space mapping are further reflected in the Raman intensity and linewidth distributions. Sample A exhibits a substantially stronger Ge-Ge Raman response than sample B, with an average integrated peak area of $15472 \pm 769$ compared with $6708 \pm 775$~arbitrary units (a.u.), respectively (Fig.~\ref{fig:raman_2d}(d,h)). At the same time, the broader linewidth and larger spatial variation of the Ge-Ge mode observed for sample B (Fig.~\ref{fig:raman_2d}(c,g)) indicate enhanced phonon scattering arising from increased crystalline disorder \cite{bouthillier2020}. These observations are fully consistent with the broader diffraction peaks in the $\omega$-2$\theta$ scans, the diffuse reciprocal-lattice intensity in the reciprocal-space maps and the increased mosaicity revealed by the pole-figure measurements. Taken together, the diffraction and Raman measurements provide a consistent picture of the structural evolution of the GeSn epilayers. Although both films are essentially fully relaxed and possess comparable average lattice parameters and alloy compositions, increasing the thickness from 70 to 140 nm promotes defect-mediated structural disorder, resulting in reduced crystalline coherence and enhanced phonon damping. The relatively small spatial variations observed across the mapped $20 \times 20~\mu\mathrm{m}^{2}$ regions nevertheless indicate good lateral uniformity of the Ge-Ge vibrational response, despite the localized surface precipitates observed in the thicker film by SEM (which are discussed in detail in the subsequent section).

\begin{figure}[!htb]
	\centering
	\includegraphics[width=0.9\linewidth]{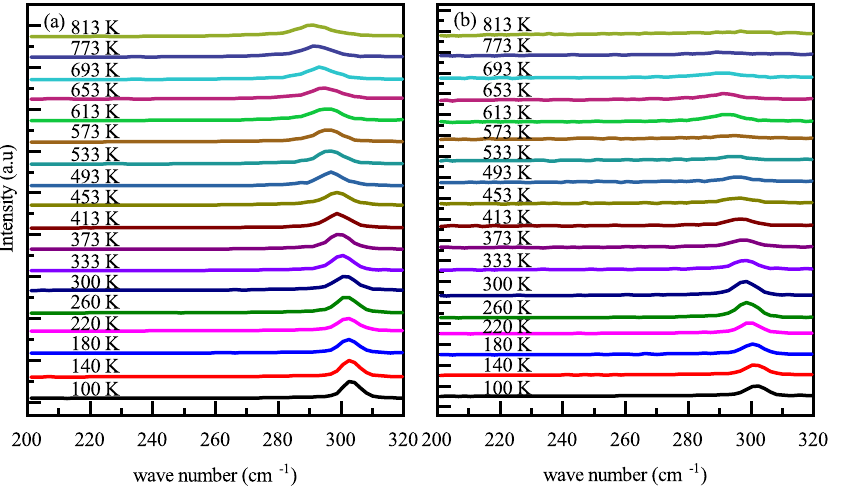}
	\caption{Temperature-dependent Raman spectra of the Ge-Ge longitudinal optical phonon region recorded during heating from 100 to 813~K in 40~K steps for (a)~sample~A (70~nm) and (b)~sample~B (140~nm). Spectra are offset vertically for clarity. The thermal cycle simultaneously probes anharmonic phonon evolution and serves as a controlled annealing protocol for the post-cycling morphological analysis.}
	\label{fig:TD_raman}
\end{figure}

Temperature-dependent Raman spectra of the Ge-Ge longitudinal optical (LO) phonon mode were acquired during heating from 100 to 813 K in steps of 40 K (Fig.~\ref{fig:TD_raman}). In both samples, the Ge-Ge phonon progressively redshifts and broadens with increasing temperature, consistent with lattice thermal expansion and anharmonic phonon scattering reported for GeSn alloys~\cite{liu2020,corley2024}. Despite this common behaviour, the thermal evolution differs markedly between the two epilayers. The Ge-Ge mode of sample A remains well defined throughout the entire temperature range, retaining a strong Raman response even at 813 K (Fig.~\ref{fig:TD_raman}(a)). In contrast, sample B exhibits a weaker and broader Ge-Ge mode from the outset, followed by a pronounced reduction in Raman intensity at elevated temperatures, such that the phonon feature becomes poorly resolved at 813 K (Fig.~\ref{fig:TD_raman}(b)). These observations indicate that the thicker GeSn epilayer is substantially more susceptible to thermally induced structural degradation, consistent with its lower crystalline quality established by the diffraction and room-temperature Raman measurements.

\begin{figure}[!htb]
	\centering
	\includegraphics[width=0.55\linewidth]{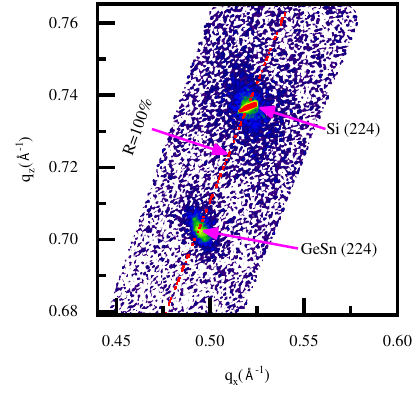}
	\caption{Asymmetric (224) reciprocal-space map of sample A (70 nm) acquired after temperature-dependent Raman measurements up to 813 K. The corresponding GeSn peak was not observed for sample B (140 nm) under identical post-cycling conditions.}
	\label{fig:post_raman_rsm}
\end{figure}

To evaluate the crystalline state following thermal cycling, HRXRD-RSM was performed after cooling the samples to room temperature. For sample A, the GeSn epilayer reflection remained clearly observable (Fig.~\ref{fig:post_raman_rsm}), yielding in-plane and out-of-plane lattice parameters of $a_{\parallel}=5.683$~\AA\ and $a_{\perp}=5.685$~\AA, corresponding to a substitutional Sn concentration of 3.15 at.\% in an essentially fully relaxed state (residual strain of $-0.02\%$). Compared with the as-grown Sn content of approximately 4.5 at.\%, this reduction indicates the loss of substitutional Sn from the coherent GeSn lattice, consistent with thermally induced Sn precipitation from the metastable alloy. In contrast, no distinct GeSn epilayer reflection was detected for sample B under identical measurement conditions, indicating that the coherent GeSn layer had largely lost its crystallinity during thermal cycling. This pronounced structural degradation is consistent with the severe attenuation of the Ge--Ge Raman mode and the extensive surface segregation observed by SEM. Taken together, these results demonstrate that the thicker GeSn epilayer exhibits substantially lower structural stability during high-temperature annealing, reflecting the greater defect density established during epitaxial growth.

\begin{figure}[!htb]
	\centering
	\includegraphics[width=0.65\linewidth]{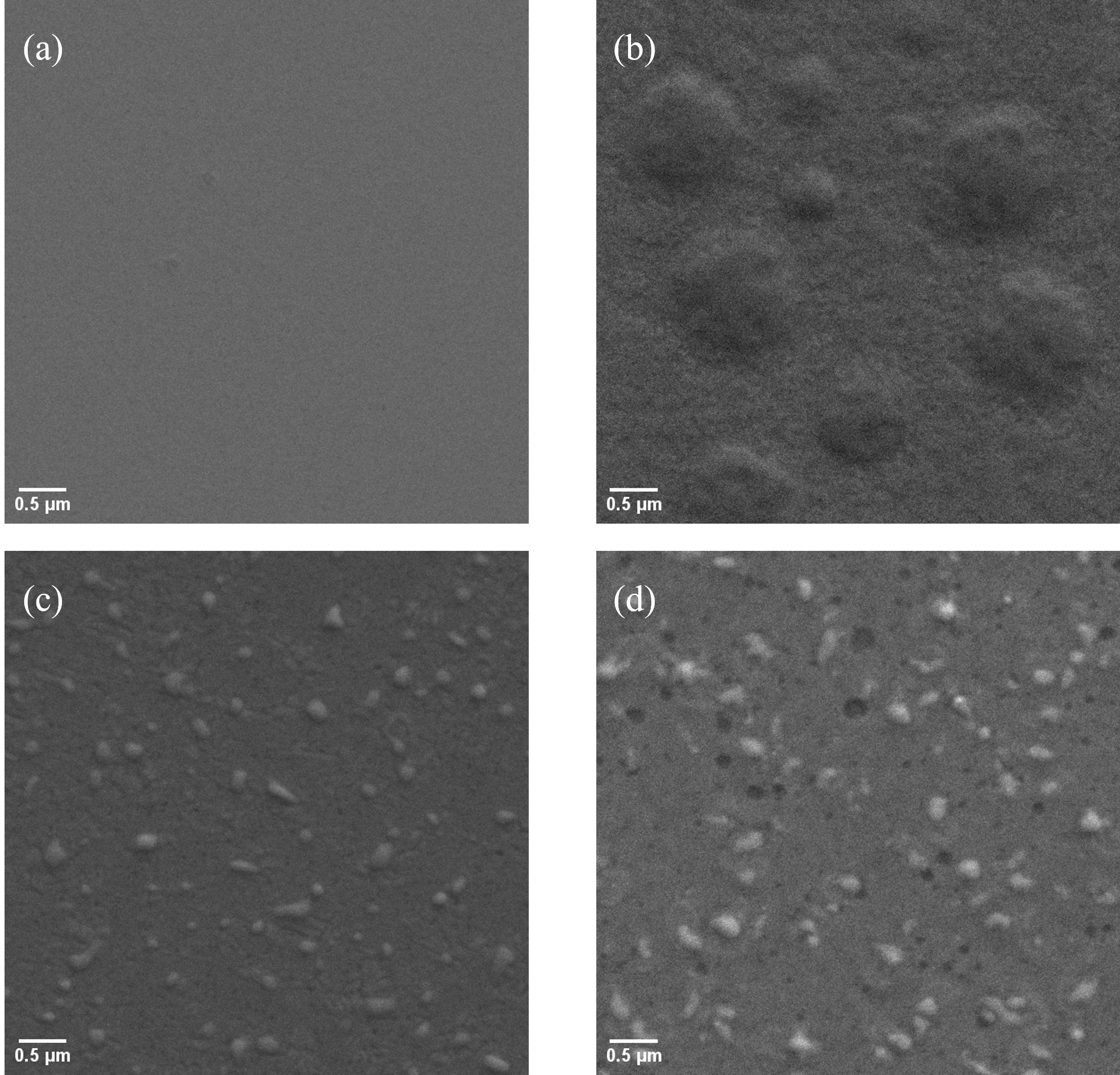}
	\caption{Scanning electron microscopy (SEM) micrographs showing the surface morphology of the samples at $\times$10{,}000 magnification: (a) as-deposited sample A (70 nm), (b) sample A after temperature-dependent Raman measurements, (c) as-deposited sample B (140 nm), and (d) sample B after temperature-dependent Raman measurements.}
	\label{fig:SEM}
\end{figure}

The structural evolution during thermal cycling is further reflected in the surface morphology (Fig.~\ref{fig:SEM}). Before heating, sample A exhibits a smooth and featureless surface, whereas sample B already contains a high density of nanoscale surface precipitates. Because the Sn concentration of both epilayers exceeds the equilibrium solid solubility of Sn in Ge, the alloy is intrinsically metastable and maintained by non-equilibrium epitaxial growth \cite{zhang2017,groiss2017}. The presence of surface precipitates in the thicker film is therefore consistent with an increased tendency for Sn segregation, facilitated by its higher defect density and the larger volume of the film acting as a source for diffusing Sn atoms~\cite{comrie2016}.

After heating to 813 K, the morphological differences become more pronounced. Sample A develops only shallow surface undulations while remaining free from discrete surface islands. By contrast, sample B exhibits a clear increase in both the size and density of surface precipitates together with the appearance of pit-like features, indicating enhanced surface segregation and morphological instability during thermal treatment~\cite{comrie2016,zhang2017}. The more extensive surface evolution of sample B closely correlates with the pronounced degradation of its Ge-Ge Raman response, suggesting that both observations originate from the same defect-mediated thermal evolution.

Taken together, the temperature-dependent Raman and SEM measurements demonstrate that the thermal stability of direct GeSn/Si(001) heterostructures is strongly influenced by the initial crystalline quality. Although the co-variation of thickness, Sn composition, and residual strain in our two samples prevents a complete decoupling of these parameters, the significantly higher defect density in the thicker film is expected to dominate the thermal response by facilitating pipe diffusion of Sn~\cite{nicolas2020}. Further studies involving intermediate film thicknesses, varied defect densities at constant thickness, or detailed composition mapping (e.g., energy-dispersive X-ray spectroscopy (EDS) or X-ray photoelectron spectroscopy (XPS)) are needed to fully isolate these effects.

\section{Conclusion}
Epitaxial Ge$_{1-x}$Sn$_x$ films containing approximately $\sim$~4.5 at.\% Sn were grown directly on Si(001) by low-temperature molecular beam epitaxy with thicknesses of 70 and 140 nm, and their structural quality and thermal stability were systematically investigated. Reciprocal-space mapping demonstrates that both epilayers are fully relaxed, with the thicker film exhibiting a slightly higher Sn composition ($5.0\%$ vs $4.5\%$) and a small residual in-plane tensile strain ($+0.09\%$) at room temperature. Nevertheless, increasing the film thickness results in a progressive degradation of the crystalline quality, manifested by broader HRXRD peaks, diffuse reciprocal-space intensity, and a weaker, less uniform Ge-Ge Raman response, indicating that defect accumulation rather than the macroscopic strain state dominates the thicker epilayer.

Temperature-dependent Raman spectroscopy and SEM characterization further reveal that the thermal stability is strongly correlated with the initial crystalline quality. While the 70 nm film remains stable up to 813 K, the 140 nm film undergoes substantial structural and morphological degradation. Although composition and strain differences co-vary with thickness, the accelerated degradation in the thicker film is highly consistent with defect-mediated Sn segregation and diffusion. These insights highlight the importance of controlling the initial defect structure to optimize the reliability of GeSn-based optoelectronic devices.

\section*{Acknowledgments}
The authors acknowledge the Central Instrumentation Facility, Indian Institute of Technology Gandhinagar, for providing access to the HRXRD and Raman spectroscopy facilities. The authors also thank the IIT Bombay Nanofabrication Facility (IITBNF) for support during sample preparation. TVV and KRK gratefully acknowledge funding support from the Science and Engineering Research Board (SERB), Department of Science and Technology (DST), Government of India, under grant no. EEQ/2022/000354.

\section*{Data availability}
The data that support the findings of this study are available from the corresponding author upon reasonable request.

\bibliographystyle{unsrt}
\bibliography{References}

\end{document}